\documentclass[aps,preprints,preprintnumbers,showpacs,superscriptaddress,groupedaddress]{revtex4-1}  
\usepackage{graphicx}  
\usepackage{dcolumn}   
\usepackage{bm}        
\usepackage{amssymb}   
\usepackage{color}
\usepackage{CJK,upgreek,fancyhdr}

\usepackage{psfrag}
\usepackage{epsfig}
\usepackage{epsf}
\usepackage{float}
\usepackage{slashed}

\newcommand{\beq}{\begin{equation}}
\newcommand{\eeq}{\end{equation}}
\newcommand{\beqa}{\begin{eqnarray}}
\newcommand{\eeqa}{\end{eqnarray}}

\begin{document}
\begin{CJK*}{UTF8}{gbsn}

\title{On the chiral covariant approach to $\rho\rho$ scattering}

\author{Li-Sheng~Geng(耿立升)}
\email[]{Email: lisheng.geng@buaa.edu.cn}
\affiliation{School of Physics and
Nuclear Energy Engineering \& International Research Center for Nuclei and Particles in the Cosmos \& Beijing Key Laboratory of Advanced Nuclear Materials and Physics, Beihang University, Beijing 100191, China}

\author{Raquel Molina}

\affiliation{Physics Department, The George Washington University, Washington, DC 20052, USA }

\author{Eulogio Oset}

\affiliation{Departamento de
F\'{\i}sica Te\'orica and IFIC, Centro Mixto Universidad de
Valencia-CSIC Institutos de Investigaci\'on de Paterna, Aptdo.
22085, 46071 Valencia, Spain}

\date{\today}

\begin{abstract}
We examine in detail a recent work (D.~G\"ulmez, U.-G.~Mei\ss ner and J.~A.~Oller, Eur. Phys. J. C 77:460 (2017)), where improvements to make  $\rho\rho$ scattering relativistically covariant are made. The paper has the remarkable
conclusion that the $J=2$ state disappears with a potential which is much more attractive than for $J=0$, where a bound state is found. We trace this abnormal conclusion
to the fact that an ``on-shell'' factorization of the potential is done in a region where this potential is singular and develops a large discontinuous and unphysical imaginary part. A method is developed, evaluating the loops with full $\rho$ propagators, and we show that they do not develop singularities and do not have an imaginary part below threshold. With this result for the loops we define an effective potential, which when used with the Bethe-Salpeter equation provides a state with $J=2$ around the energy of the $f_2(1270)$. In addition, the coupling of the state to $\rho\rho$ is evaluated and we find that this coupling and the $T$ matrix around the
energy of the bound state are remarkably similar to those obtained with a drastic approximation used previously, in which the $q^2$ terms of the propagators of the exchanged $\rho$ mesons are dropped, once the cut-off in the $\rho\rho$ loop function is tuned to reproduce the bound state at the same energy.
\end{abstract}

\maketitle

\section{Introduction} 
The chiral unitary approach, combining the dynamical features of chiral Lagrangians and unitarity in coupled channels, has allowed much progress
in the meson-meson~\cite{npa, kaiser, locher, juannito} and meson-baryon interactions~\cite{weise,angels, ollerulf, carmen, hyodo} (see review paper~\cite{review}). One step forward in this direction was the extension of the approach to study the interaction of vector mesons among themselves. The first such work studied
the $\rho\rho$ interaction~\cite{raquel}, which was found to be attractive in the isospin $I=0$ and spin $J=0,2$ channels. The strength of the interaction in the $J=2$ channel was found 
more than twice as big as that of the $J=0$ channel. In both cases it was sufficient to produce bound states.  The one with $J=0$ was associated to the $f_0(1370)$ and 
the one with $J=2$ to the $f_2(1270)$ states. The work was generalized to the SU(3) sector~\cite{geng} and more resonant states were found that could be associated
with known states. 

In Refs.~\cite{raquel,geng} the parameters of the loop function were fine tuned. With natural values of the parameters in order to find the binding at
the experimental energies,  the couplings of the resonances to different channels were extracted. These couplings were then used to study
radiative decays~\cite{nagahiro} and other decays~\cite{minirev}, and in all cases consistency with experiment was found.

References~\cite{raquel,geng} relied upon an approximation of neglecting the three momenta of the vector mesons with respect to their mass. This approximation was questioned in a recent work~\cite{ollernew} where improvements were made to give a fully relativistic approach. The authors found that in the $\rho\rho$ interaction
the $I=J=0$ state, the $f_0(1370)$, was obtained, very close to the result of  Ref.~\cite{raquel}, but the $f_2(1270)$ did not appear. This is certainly surprising because if 
the $f_0(1370)$ appears bound, the $f_2(1270)$, where the interaction is also attractive and with a strength more than double the one in the $I=J=0$ sector, should also appear as a bound state. 
In the present paper we show the
reasons for the findings of Ref. [15], stemming from an unjustified on-shell factorization of the potential, which renders it singular. The singularity does not appear in a proper loop function, which we evaluate
here. We propose a different method based on the results for the loop function without factorizing the propagators and show that in that case the $I=0,J=2$ channel generates a bound state, more bound than the $I=J=0$ state. The other important finding here is that if the parameters to regularize the loop are tuned to obtain the $f_2(1270)$ bound at the
experimental energy, the coupling of the state to $\rho\rho$ is very close  to the one obtained with the non-relativistic approach of Ref.~\cite{raquel}. It is well known that for composite states, and the case of a small binding, the coupling is only tied to the binding energy \cite{weinberg,kalash,gamer}. In the case of the $f_2(1270)$ the binding is 270 MeV with respect to the nominal two $\rho$ masses. Yet, this number is misleading because the $\rho$ has a width of 150 MeV and with two $\rho$ mesons their mass components go more than 300 MeV below the nominal mass and the binding is not as extreme as it  seems. From this perspective it is not so surprising  that we find the couplings so similar in different approaches.

The claim of the $f_2(1270)$ as a dynamically generated resonance from the $\rho\rho$ interaction seems at odds with a widespread belief that it
actually belongs to a $p$-wave nonet of $q\bar{q}$ states \cite{klempt,crede}. Yet, the fact remains that the molecular picture has successfully undergone  far more tests than the quark model, comparing predictions with practically all observables related to the resonance (see detailed discussions in the introduction of Refs. \cite{xiephoto1,xiephoto2}). The developments of the present work in Section 4 will further reinforce this picture.

\section{Summary of the $\rho\rho$ interaction}
\begin{figure}
  \centering
  \includegraphics[width=0.4\textwidth]{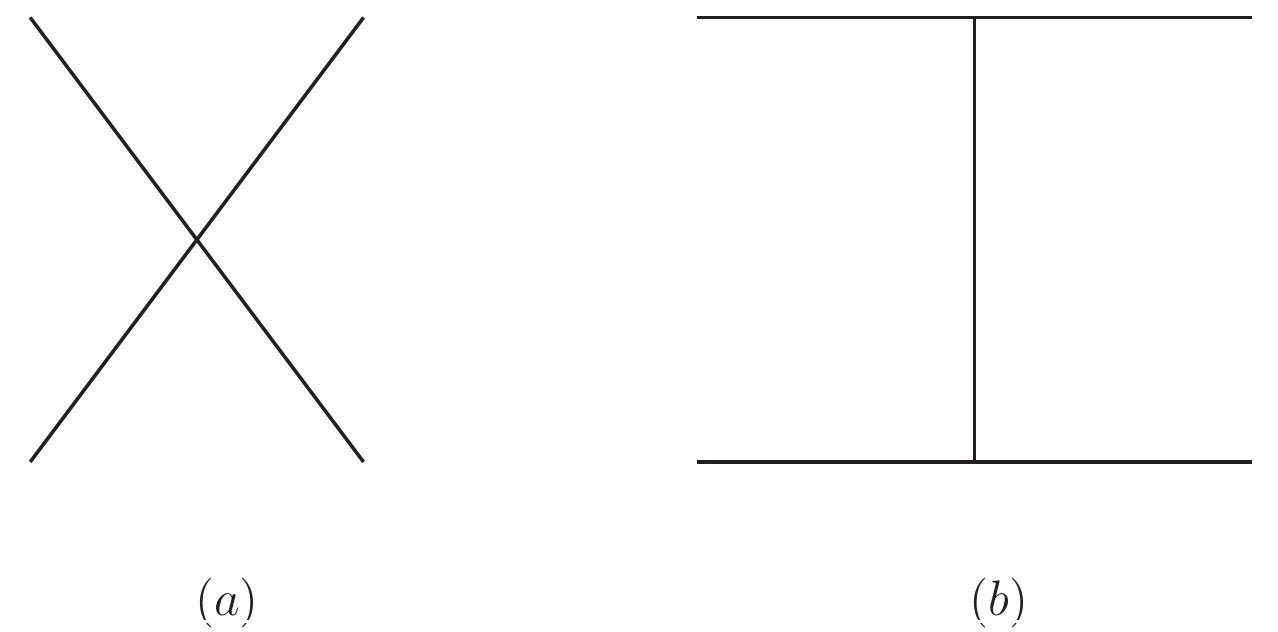} 
  \caption{Terms in the $\rho\rho$ interaction: (a) contact term; (b) $\rho$ exchange term.}\label{fig:1}
\end{figure}

In Ref. \cite{raquel} the local hidden gauge approach~\cite{hidden1,hidden2,hidden4} was used to generate the $\rho\rho$ interaction. The formalism leads to two 
terms, a contact term and a $\rho$ exchange term, which are depicted in Fig.~\ref{fig:1}. 

In Ref. \cite{raquel}, the $\rho$ exchange propagator was taken as $1/(-M_\rho^2)$, where the $q^2$ dependence of the propagator was removed. This is done in analogy to the
more general case in pseudoscalar interactions where the standard lowest order chiral Lagrangians can be obtained from the local hidden gauge approach, exchanging
vector mesons and removing the $q^2$ term in the propagator. There is another approximation made in Ref.~\cite{raquel}, since the three body vertex
$\rho\rho\rho$ contains six terms and only the two leading terms were kept, neglecting terms that go like $p_\rho/M_\rho$. This is improved in Ref.~\cite{ollernew}. With these approximations the interaction obtained in Ref.~\cite{raquel} is given in Table \ref{tab:1}, with $g=M_V/(2f)$, $M_V$ the vector mass and $f$ the pion decay constant $f=93$ MeV. 
\begin{table*}
\centering
 \caption{Potential $V$ for the scalar and tensor channels with $I=0$.}
   \begin{tabular}{cccccc}
  \hline
  \hline
 I & J& Contact & Exchange & Total at threshold $[I^G(J^{PC})]$\\
 \hline
 $0$ & $0$ & $8g^2$ & $-8g^2\left(\frac{3s}{4 M_\rho^2}-1\right)$ &$ -8g^2 [0^+(0^{++})]$\\
 $0$ & $2$ & $-4 g^2$ & $-8g^2\left(\frac{3s}{4M_\rho^2}-1\right)$ &$ -20g^2 [0^+(2^{++})]$\\
  \hline
  \hline
\end{tabular}\label{tab:1}
 \end{table*}
 
\begin{figure}
  \centering
  \includegraphics[width=0.7\textwidth]{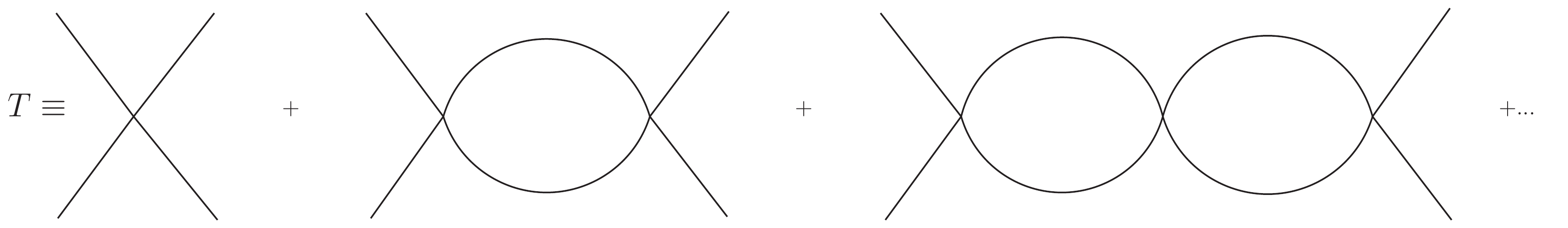} 
  \caption{Diagrammatic representation of the $\rho\rho$ scattering matrix.}\label{fig:2}
\end{figure}

One can see that the attraction in the case of $J=2$ is much bigger than in $J=0$. With the interaction in Table \ref{tab:1} one can solve the Bethe-Salpeter equation (BS),
\begin{equation}\label{eq:BS}
T=[1-VG]^{-1}V,
\end{equation}
where $G$ is the loop function of two $\rho$ meson propagators. Since the interaction has been reduced to a constant (independent of momentum transfer) for each value of $s$, the square of the total mass in the $\rho\rho$ rest frame, the amplitude $T$ in Eq.~(\ref{eq:BS}) is summing the diagrams of Fig.~\ref{fig:2}, and $G$ is given in the cut-off regularization by
\begin{equation}\label{eq:6.1}
G=\int\limits_{|\vec{q}|\le q_\mathrm{max}}\frac{d^3q}{(2\pi)^3}\frac{\omega_1+\omega_2}{2\omega_1\omega_2[P^{0\,2}-(\omega_1+\omega_2)^2+i\epsilon]}
\end{equation}
where $q_\mathrm{max}$ stands for the cutoff, $(P^0)^2=s$ and $\omega_i=\sqrt{\vec{q}\,^2+M_\rho^2}$. However, in the case of the $\rho$, which has a large width, one cannot neglect its mass distribution. This is very important and was taken into account in Ref.~\cite{raquel} by making a convolution of $G$ over the mass distribution of the two $\rho$ mesons as follows:

\begin{equation}\label{eq:G}
\tilde{G}(s)= \frac{1}{N^2}\int^{(M_\rho+2\Gamma_\rho)^2}_{(M_\rho-2\Gamma_\rho)^2}d\tilde{m}^2_1(-\frac{1}{\pi}) {\cal I}m\frac{1}{\tilde{m}^2_1-M^2_\rho+i\Gamma\tilde{m}_1}\times\int^{(M_\rho+2\Gamma_\rho)^2}_{(M_\rho-2\Gamma_\rho)^2}d\tilde{m}^2_2(-\frac{1}{\pi}) {\cal I}m\frac{1}{\tilde{m}^2_2-M^2_\rho+i\Gamma\tilde{m}_2} G(s,\tilde{m}^2_1,\tilde{m}^2_2)\ ,
\label{Gconvolution}
\end{equation}
with
\begin{equation}\label{eq:N}
N=\int^{(M_\rho+2\Gamma_\rho)^2}_{(M_\rho-2\Gamma_\rho)^2}d\tilde{m}^2_1(-\frac{1}{\pi}){\cal I}m\frac{1}{\tilde{m}^2_1-M^2_\rho+i\Gamma\tilde{m}_1}\ ,
\label{Norm}
\end{equation}
where $M_\rho=770$ MeV, $\Gamma_\rho=146.2$ MeV and for $\Gamma\equiv\Gamma(\tilde{m})$ we take the $\rho$ width for the decay into  pions in the $p$-wave
\begin{equation}\label{eq:Gamma}
\Gamma(\tilde{m})=\Gamma_\rho (\frac{\tilde{m}^2-4m^2_\pi}{M^2_\rho-4m^2_\pi})^{3/2}\theta(\tilde{m}-2m_\pi).
\label{gamma}
\end{equation}

The use of this $\tilde{G}$ function gives a width to the bound states obtained from the $\rho\rightarrow\pi\pi$ decay. In addition, box diagrams with four intermediate $\pi$ mesons were also considered in Ref.~\cite{raquel}, which account for the $\pi\pi$ decay channel of the states obtained. This channel is not a matter of concern in Ref.~\cite{ollernew} and we shall not discuss it here.

One should bear in mind that we are working with an effective theory, which is not renormalizable. This is generally the case in all 
effective theories, in particular, chiral perturbation theory \cite{weinbergchi,gasser,scherer}. Despite this, the loops are well defined, with prescription given for their regularization and introducing appropriate counterterms, 
and the theory is remarkably successful at low energies. One can 
proceed in a similar way
in the case of the local hidden gauge Lagrangians,  from which the chiral Lagrangians can actually be obtained \cite{derafael}. Equation (\ref{eq:BS}), from which a unitary amplitude is 
constructed, can be obtained imposing the unitary constraint, $\mathrm{Im} t=t^*\sigma t$, (with $\sigma$ the phase space for the intermediate state proportional to the momentum). The latter equation can be recast as $\mathrm{Im} t^{-1}=-\sigma$, which allows use of a
dispersion relation for $t^{-1}$ that is made convergent with a subtraction constant for the $s$-wave, which we study here. This was done in Refs.~\cite{chew,nsd,sigma}, and is the base of the chiral unitary approach \cite{npa,kaiser,locher,juannito,weise,angels,ollerulf,carmen,hyodo}. In Ref.~\cite{ollerulf}, the equivalence of the use of the dispersion relation method and the loop regularization with a cut-off was also established, and this latter method is often used in the unitary approach of effective theories \cite{review}.

Equation (\ref{eq:BS}) is generally referred to as the Bethe-Salpeter equation \cite{bethe}. This is because relativistic propagators are used for the propagation of the intermediate particles and a $d^4q$ integral is made in the 
integral equation (unlike the $d^3q$ integral of the Lippmann Schwinger equation). However, a factorization of the kernel (potential) has been done to arrive at Eq. (\ref{eq:BS}), which more properly should be called the on-shell factorized BS equation. The on-shell factorization is justified when one can neglect the contribution of the left hand cut in the dispersion relation discussed above \cite{ollerulf} or it is quite energy independent, in which case it can be reabsorbed by means of subtraction constants in the dispersion integral, which are finally obtained by fitting to some data \cite{nsd}. Note 
that the left hand cut for equal mass particles goes from $s=-\infty$ to $s=0$, still far away from the $f_2(1270)$ mass in the present problem.
 
 Concerning the kernel as being just the $\rho$ exchange, the growing energy dependence of the interaction (see Table \ref{tab:1}) has been used as one element to justify the introduction of Regge phenomena \cite{regge,regge2,adam}. The $\rho$ exchange in our approach would then be substituted by a Regge trajectory. The energy dependence for the case of $J=2$ can be seen in Table \ref{tab:1}, and one should note that, apart from the linear term in $s$, there is a constant term of strength $-12g^2$, which is more important than the $s$ dependent term around $\sqrt{s}=1270$ MeV. Yet, it would be interesting to see what differences can come from the use of the plain $\rho$ exchange or the full $\rho$ trajectory. Such a 
 test has already been done in the study of the photoproduction of the $f_2(1270)$. For the $\rho$ trajectory, Refs.~\cite{barnes,navelet,guidal,kochelev,adam2} were considered, but using a constant phase which is favored by the CLAS data \cite{clas}. The conclusion  was that both approaches gave similar results, using moderate flexibility in the parameters of the models compatible with the phenomenology of other processes.

\section{Beyond the static $\rho$ exchange with on-shell factorization}
The novelty in Ref.~\cite{ollernew}, which is the reason for the disappearance of the tensor state, stems from keeping the $q^2$ dependence in the $\rho$ exchange potential in Fig. \ref{fig:1}(b). To show that, one can still use the potential in Table \ref{tab:1}, since the relativistic improvements on the vertices have nothing to do with this problem. The $\rho$ propagator in Fig.~\ref{fig:1}(b) gives, in the notation $p_1+p_2\rightarrow p_3+p_4$,
\begin{equation}
D(\rho)=\frac{1}{q^2-M_\rho^2+i\epsilon}=\frac{1}{(p_1-p_3)^2-M_\rho^2+i\epsilon}=\frac{1}{-2\vec{p}\,^2(1-\cos\theta)-M_\rho^2+i\epsilon}
\end{equation}
where we have taken $\vec{p}_1=p \hat{u}_z$, $\vec{p}_2=-\vec{p}_1$, and as in Ref.~\cite{ollernew} we have taken $q^0=0$. This corresponds to the on-shell factorization, where the interaction $V$ is taken for an on-shell situation. The next assumption in the on-shell factorization in Ref.~\cite{ollernew} is that $p_i^2=M_\rho^2$. Thus 
$p^2=\left(\frac{E}{2}\right)^2-M_\rho^2$ and hence, $p^2$ becomes negative for bound states, $E=\sqrt{s}<2 M_\rho$. Here is where the problem begins, because the $\rho$ exchange develops a singularity. However, we can already advance that this singularity never appears in the loops of the Bethe-Salpeter equation
of Fig.~\ref{fig:2} when the  $\rho \rho \rho \rho$ vertex is substituted by the $\rho$ exchange diagram of Fig.~\ref{fig:1}(b). Continuing with the derivation, we project the $\rho$-exchange in $s$-wave as done in Ref.~\cite{ollernew} and obtain
\begin{equation}\label{eq:8.2}
D_\rho(s\mathrm{-wave})=-\frac{1}{4 p^2}\log\left(\frac{4p^2+M_\rho^2}{M_\rho^2}+i\epsilon\right).
\end{equation}
We can see that when $4p^2+M_\rho^2\equiv s-4M_\rho^2+M_\rho^2=0$, this has a singularity, and the on-shell factorized potential becomes infinite at $s=3 M_\rho^2$. In addition, for $s<3 M_\rho^2$, $D_\rho(s-\mathrm{wave})$ develops an imaginary part.

\begin{figure}
  \centering
  \includegraphics[width=0.32\textwidth]{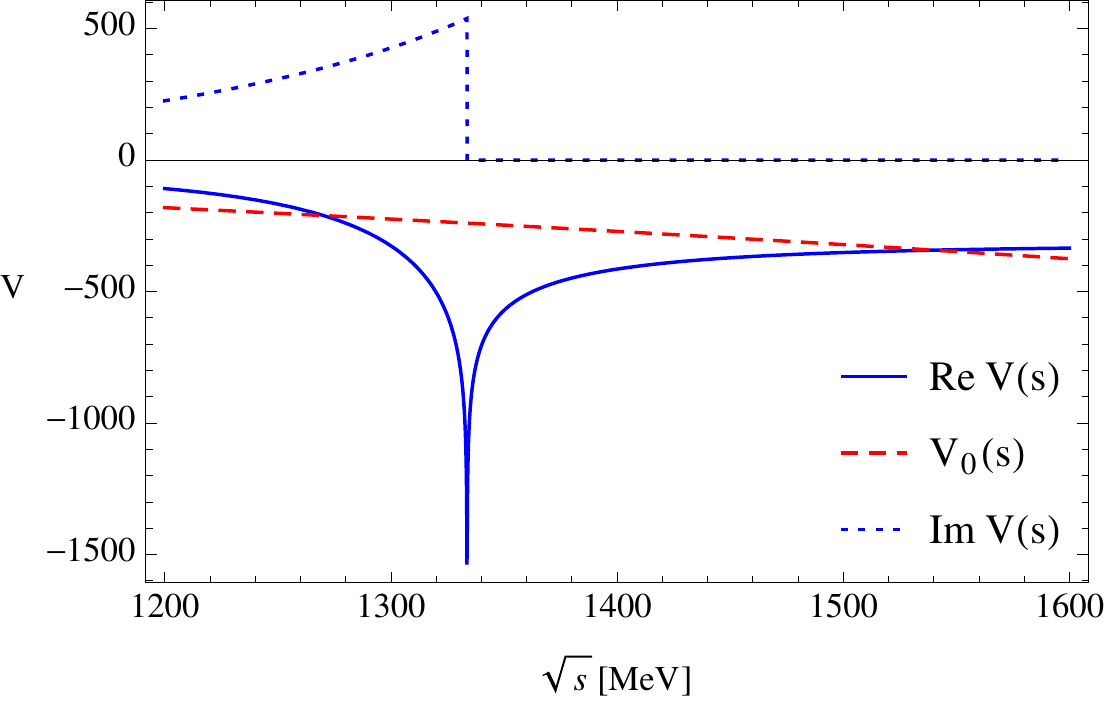} 
  \caption{Dashed line: $V_0=V_c+V_\mathrm{ex}$ from Ref.~\cite{raquel}. Solid line: Re $V(s)$ of Eq.~(\ref{eq:8.1}). Dotted line: Im $V(s)$ of Eq.~(\ref{eq:8.1})}\label{fig:3}
\end{figure}

In Fig. ~\ref{fig:3}, we plot the new potential
\begin{equation}\label{eq:8.1}
V(s)=V_c+V_\mathrm{ex}D_\rho(s-\mathrm{wave})(-M_\rho^2)
\end{equation}
with $V_c$ and $V_\mathrm{ex}$ from Table \ref{tab:1}, where we have replaced $\frac{1}{-M_\rho^2}$ by $D_\rho(s-\mathrm{wave})$ in the $V_\mathrm{ex}$ potential of
Ref.~\cite{raquel}.

\begin{figure}
  \centering
  \includegraphics[width=0.32\textwidth]{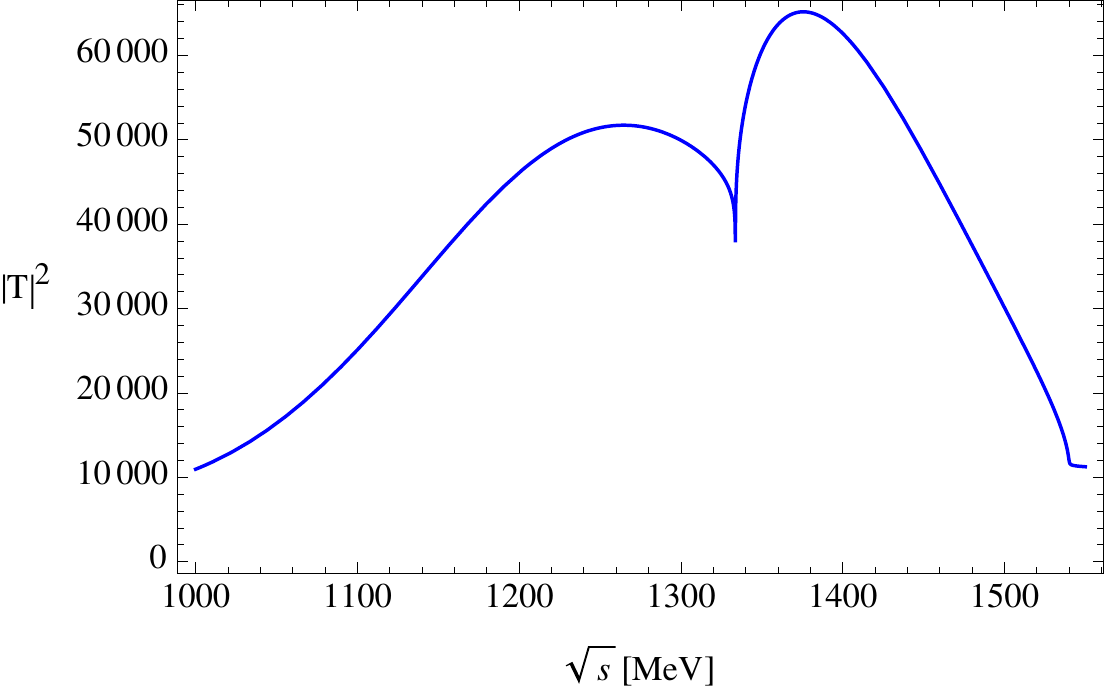} 
  \caption{The results for $|T|^2$ with the potential of Eq.~(\ref{eq:8.1}).}\label{fig:4}
\end{figure}

As we can see, the new potential of Eq.~(\ref{eq:8.1}) is remarkably similar to the one exhibited in Fig. 4 of Ref.~\cite{ollernew}. It is exactly equal to the one of Ref.~\cite{raquel}
at threshold and develops a singularity at $s=3M_\rho^2$. One can also see that the potential develops an imaginary part for $s<3M_\rho^2$, with a discontinuity at $s=3M_\rho^2$. This imaginary part is not tied to any physical process, as could be the $\rho\rho$ system decaying into $2\pi$ or $4\pi$. The singularity appears at $\sqrt{s}=1334$ MeV and, hence, one anticipates problems to get a state at 1270 MeV, as it would correspond to the
$f_2(1270)$ state. Indeed, in Fig. \ref{fig:4} we plot the result of $|T|^2$ for this potential. As we can see, this does not reflect a resonance at 1270 MeV with a width of 100 MeV as in the experiment. In this sense, the conclusion of Ref.~\cite{ollernew} that the tensor resonance $f_2(1270)$ does not appear with the potential of Eq.~(\ref{eq:8.1}) is correct. The problem is that this is a clear situation where the on shell factorization cannot be done since the ``on- shell'' potential seats on top of a singularity of the extrapolated amplitude below threshold. 

Before we proceed to perform the integration of the loop function with the full $\rho$ propagator (including also the $q^0$ dependence) let us, however, note that the singularity obtained corresponds to using a $\rho$ mass fixed  to the nominal value of 770 MeV. We next show what happens if the realistic $\rho$ mass distribution is used. For this, we again take the $D_\rho(s-\mathrm{wave})$ of Eq.~(\ref{eq:8.2}) and convolute it  with the $\rho$ mass distribution. Hence, we now use 
\begin{equation}\label{eq:11.1}
\tilde{V}(s)=V_c+V_\mathrm{ex} \tilde{D}_\rho(s-\mathrm{wave})(-M_\rho^2)
\end{equation}
\bibliography{refs}
with
\begin{equation}\label{eq:11.2}
\hat{D}_\rho=\frac{1}{N}\int^{(M_\rho+2\Gamma_\rho)^2}_{(M_\rho-2\Gamma_\rho)^2}d\tilde{m}_\rho^2\left(-\frac{1}{\pi}\right)
\mathrm{Im}\frac{1}{\tilde{m}_\rho^2-M_\rho^2+i\Gamma\tilde{m}_\rho} \left[
-\frac{1}{4 p^2}\log\left(\frac{4p^2+\tilde{m}_\rho^2}{\tilde{m}_\rho^2}+i\epsilon\right)\right]
\end{equation}
with $N$ and $\Gamma$ given by Eqs. ~(\ref{eq:N})(\ref{eq:Gamma}), and $p^2=\frac{s}{2}-\tilde{m}_\rho^2$. 

\begin{figure}
  \centering
  \includegraphics[width=0.32\textwidth]{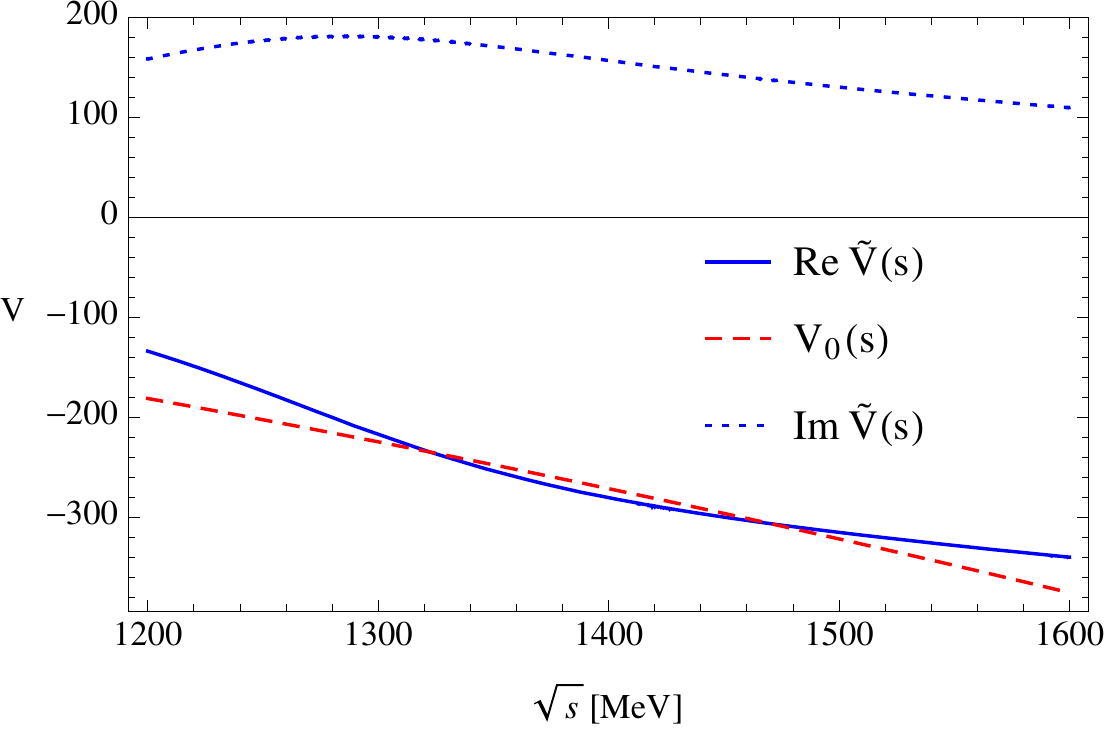} 
  \caption{The potential of Eqs.~(\ref{eq:11.1}) and (\ref{eq:11.2}).}\label{fig:5}
\end{figure}

\begin{figure}
  \centering
  \includegraphics[width=0.32\textwidth]{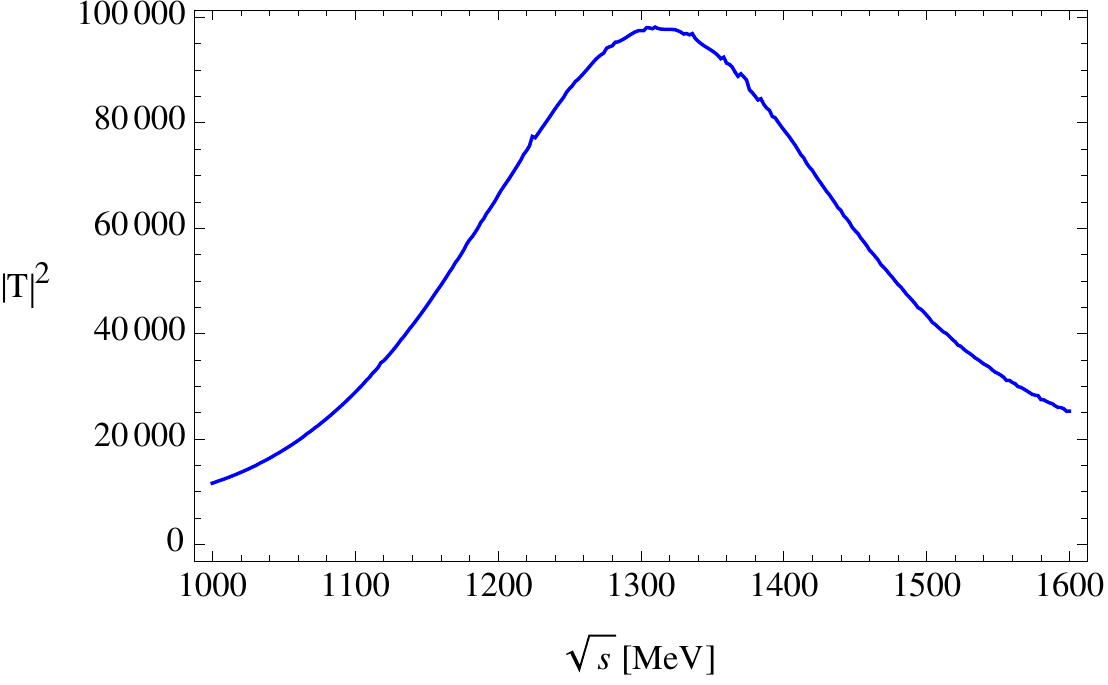} 
  \caption{$|T|^2$ from $\tilde{V}(s)$ of Eqs.~(\ref{eq:11.1}) and (\ref{eq:11.2}) and $\tilde{G}$ from Eq.~(\ref{eq:G}) via Eq.~(\ref{eq:BS}).}\label{fig:6}
\end{figure}

\begin{figure}
  \centering
  \begin{tabular}{cc}
  \includegraphics[width=0.32\textwidth]{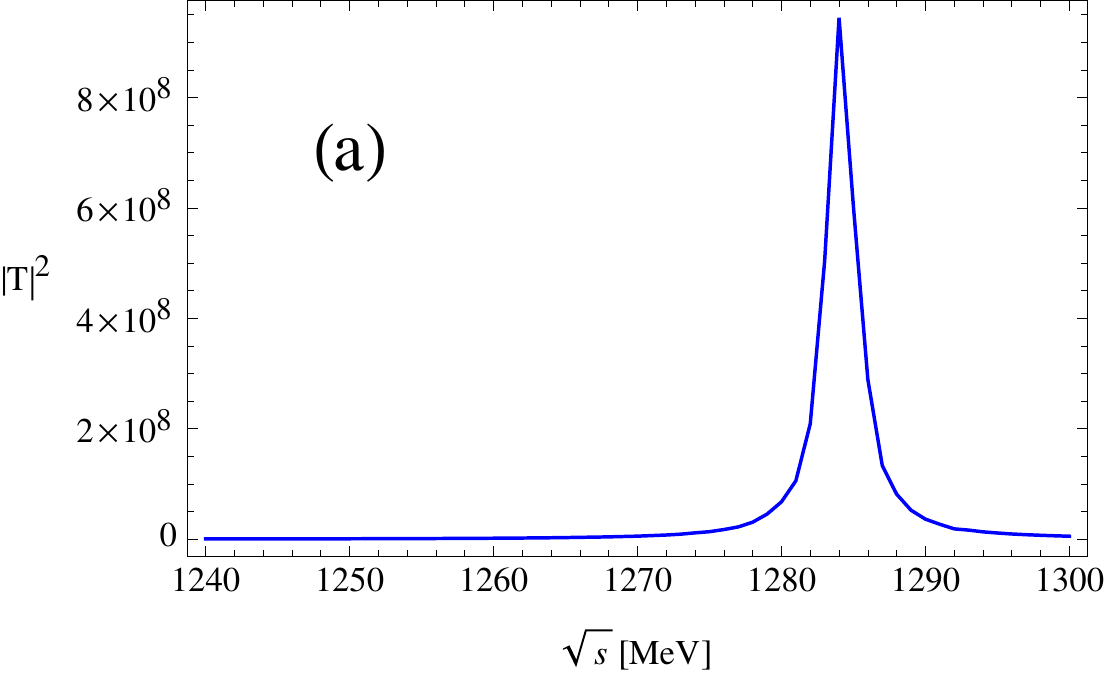} &
    \includegraphics[width=0.32\textwidth]{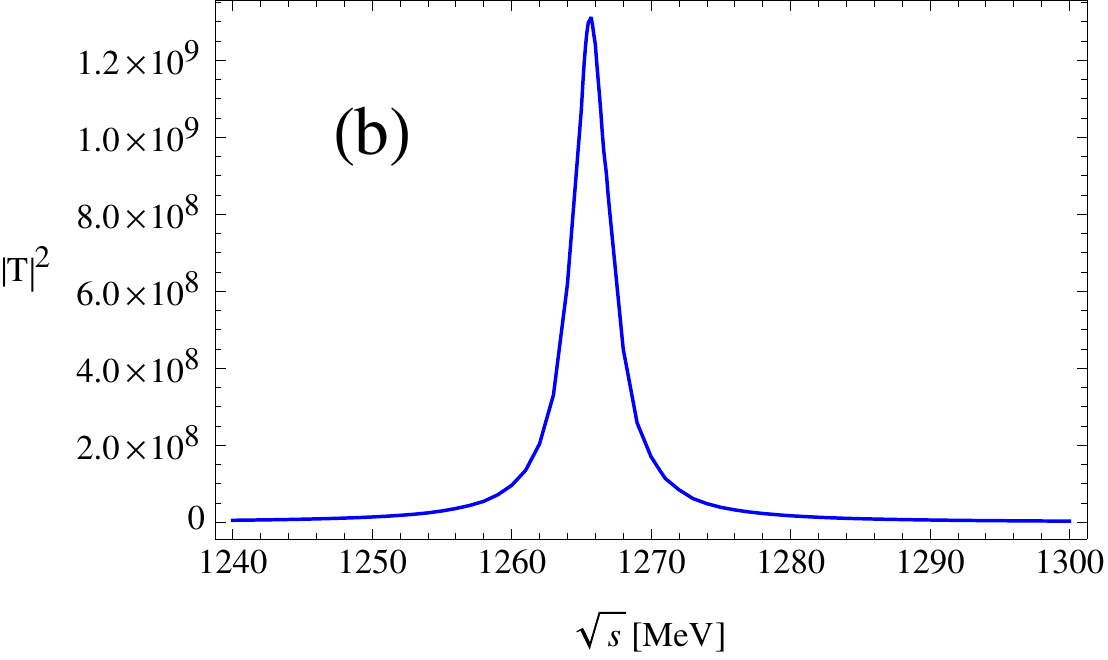} \\
    \end{tabular}
  \caption{$|T|^2$ obtained from $\mathrm{Re}\tilde{V}(s)$ (a) and from $V_c+V_\mathrm{ex}$ of Ref.~\cite{raquel} (b).}\label{fig:7}
\end{figure}

In Fig.~\ref{fig:5} we show $\tilde{V}(s)$ compared to that from Ref.~\cite{raquel}. We can see that now $\tilde{V}(s)$ does not have a singularity and $\mathrm{Re}\tilde{V}(s)$
is actually quite similar to the potential from Ref.~\cite{raquel}. In addition, the imaginary part of $\tilde{V}(s)$ no longer has a discontinuity. It is interesting to see what happens if we use the Bethe-Salpeter equation with this potential. In Fig.~\ref{fig:6}, we show $|T|^2$ evaluated with the potential $\tilde{V}(s)$ and Eq. (\ref{eq:BS}) with the same cut-off $q_\mathrm{max}=875$ MeV as in Ref.~\cite{raquel}. Using $\tilde{G}(s)$ from Eq.~(\ref{eq:G}), we get a broad bump that could be identified with a resonance with mass around 1300 MeV and $\Gamma\approx300$ MeV. So, even using the on shell approach of Ref.~\cite{ollernew}, a state with mass around 1300 MeV appears. The width, however, is not realistic, since it is  related to the imaginary part of the $D_\rho(s-\mathrm{wave})$, which is not linked to any physical channel. If we remove this spurious imaginary part, we obtain for $|T|^2$ the result shown in Fig. ~\ref{fig:7}(a), which is remarkably close to that of Ref.~\cite{raquel}, shown in Fig.~\ref{fig:7}(b). 

\begin{figure}
  \centering
  \includegraphics[width=0.9\textwidth]{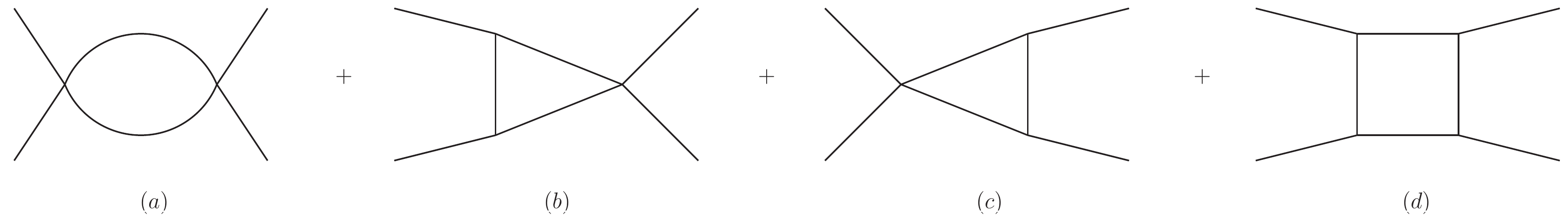} 
  \caption{Diagrams appearing at one-loop level with the contact and $\rho$ exchange terms.}\label{fig:8}
\end{figure}

\begin{figure}
  \centering
  \includegraphics[width=0.32\textwidth]{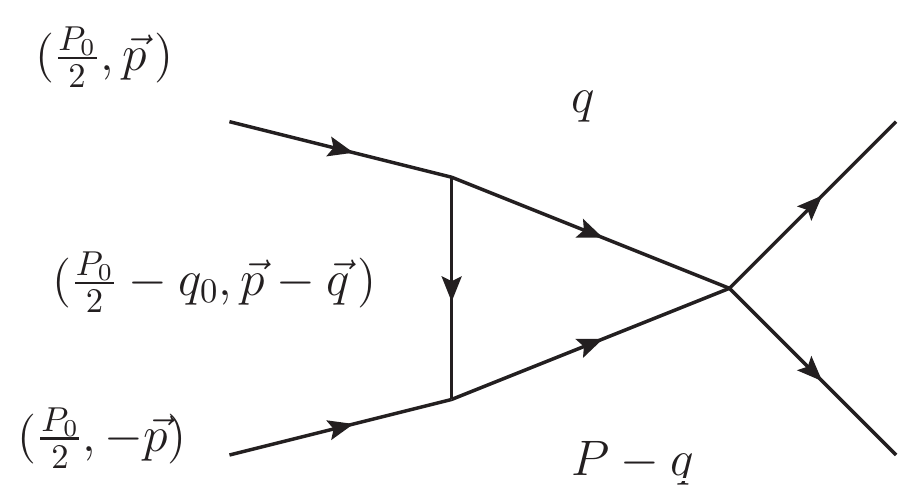} 
  \caption{Diagram of Fig.~\ref{fig:8}(b) showing explicitly the momenta of the particles.}\label{fig:9}
\end{figure}


\begin{figure}
  \centering
  \includegraphics[width=0.32\textwidth]{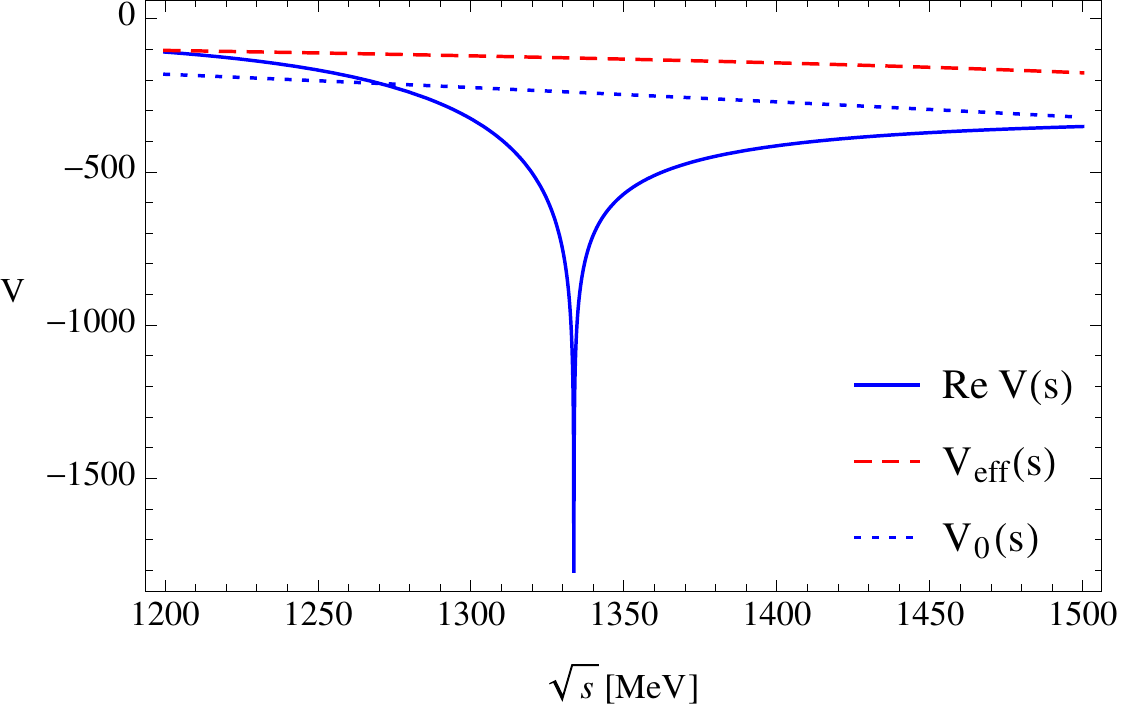} 
  \caption{Comparison of $V_\mathrm{eff}$, Re $V(s)$ and the potential from Ref.~\cite{raquel}. }\label{fig:11}
\end{figure}

\begin{figure}
  \centering
  \includegraphics[width=0.32\textwidth]{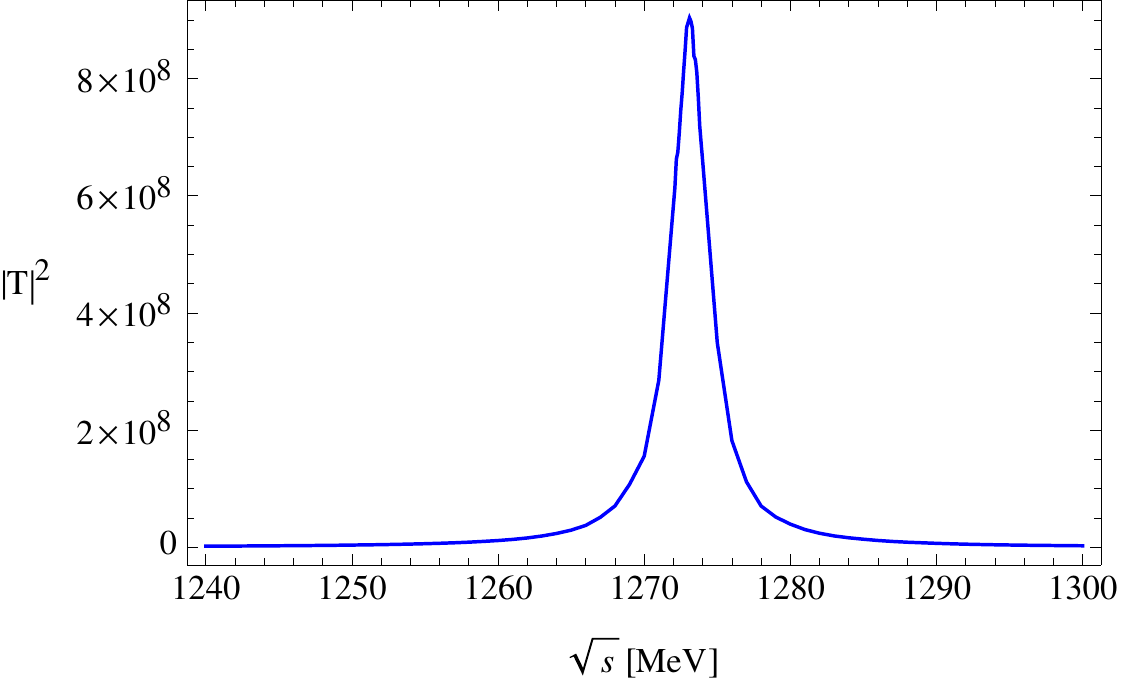} 
  \caption{$|T|^2$ evaluated with $V_\mathrm{eff}$ and $\tilde{G}$. The value of $q_\mathrm{max}$ is 1500 MeV. }\label{fig:12}
\end{figure}

\vspace{1cm}
\section{Improved calculation}
In this section we are going to evaluate explicitly the loops that would appear in the Bethe-Salpeter equation, Fig.~(\ref{fig:2}), where a contact term or the explicit $\rho$ exchange
are used as the source of interaction. We note that at the one-loop level we would have the diagrams of Fig.~\ref{fig:8}.

Next we see that around 1270 MeV we have
\begin{equation}
V_c=-4 g^2;\quad\quad V_\mathrm{ex}=-8g^2\frac{3s-4M_\rho^2}{4M_\rho^2};\quad\quad \frac{V_\mathrm{ex}}{V_c}=\frac{3s-4M_\rho^2}{2M_\rho^2}\label{eq:vs}
\end{equation}
if the exchanged $\rho$ propagator is factorized as $-\frac{1}{M_\rho^2}$. Therefore $V_\mathrm{ex}/V_c$ is of the order of two. The sum of the strength of the two middle diagrams (b) and (c) of Fig.~\ref{fig:8} will be about the same as in  Fig.~\ref{fig:8}(d), actually
even bigger when the loop is evaluated because of the reduction in the $\rho$ propagator due to the explicit consideration of the full propagator, as we shall see. We therefore concentrate on the diagram of Fig.~\ref{fig:8}(b) and evaluate it explicitly.

First, we want to see the difference between this diagram evaluated exactly and the same one when the $\rho$ propagator is replaced by $-\frac{1}{M_\rho^2}$ as in
Ref.~\cite{raquel}. For this, we neglect the vertices for the moment  and concentrate on the propagators.

In Fig. 9 we show explicitly the momenta of the variables. The loop function for this diagram considering only the propagators is given in the rest frame of the $\rho\rho$ system, $\vec{P}=0$, by
\begin{equation}\label{eq:15.1}
t=i\int\frac{d^4 q}{(2\pi)^4}\frac{1}{(\frac{P^0}{2}-q^0)^2-(\vec{p}-\vec{q})^2-M_\rho^2+i\epsilon}\frac{1}{2\omega(q)}\frac{1}{q^0-\omega(q)+i\epsilon}
\frac{1}{2\omega(q)}\frac{1}{(P^0-q^0)-\omega(q)+i\epsilon}
\end{equation}
with $\omega(q)=\sqrt{\vec{q}^2+M_\rho^2}$, where we have kept the full $\rho$ propagator for the exchanged $\rho$ (including the energy dependence). For the two intermediate $\rho$ we keep their relativistic form but keep only the positive energy part of the propagator, since they will propagate close to on-shell. 
There is practically no change from keeping the full propagators and the formulas are simplified, yet show all the analytical structure.  By analytically performing  the $q^0$ interaction in Eq.~(\ref{eq:15.1}), we obtain
\begin{equation}\label{eq:eff}
t=\int\limits_{|\vec{q}|<q_\mathrm{max}} \frac{d^3q}{(2\pi)^3}\frac{1}{2\omega(q)^2}\frac{1}{2\omega(\vec{p}-\vec{q})}\frac{1}{P^0-2\omega(q)+i\epsilon}\frac{1}{\frac{P^0}{2}-\omega(q)-\omega(\vec{p}-\vec{q})+i\epsilon}.
\end{equation}

It is interesting to look at the analytical structure of the loop. We see two cuts, the one coming from $P^0-2\omega(q)+i\epsilon$ in the denominator, which corresponds to having the two intermediate $\rho$ mesons on-shell (the two lines in the diagrams of Fig.~\ref{fig:9} cut by a vertical line) and from $\frac{P^0}{2}-\omega(q)-\omega(\vec{q}-\vec{q})+i\epsilon$ in the denominator, which accounts for a possible situation where the exchanged $\rho$ and one intermediate $\rho$ are placed on-shell. Yet, since $\omega(q)\ge M_\rho$,  for a $\rho\rho$ system below threshold, where $P^0<2M_\rho$, this term never vanishes. We can, therefore, see that the exchanged $\rho$ in the actual loops cannot produce any imaginary part, contrary to the ``on-shell'' factorization of $V_\rho(s-\mathrm{wave})$ of Eq.~(\ref{eq:8.2}), done in Ref.~\cite{ollernew}.

Performing the same calculation with $\frac{1}{-M_\rho^2}$ for the exchanged $\rho$ propagator, we obtain
\begin{equation}\label{eq:eff1}
t_f=\int\frac{d^3 q}{(2\pi)^3}\left(-\frac{1}{M_\rho^2}\right)\frac{1}{4\omega(q)^2}\frac{1}{P^0-2\omega+i\epsilon}.
\end{equation}
Comparing to $t$ in Eq.~(\ref{eq:eff}), we see that we have replaced $\frac{1}{2\omega(\vec{p}-\vec{q})}\frac{1}{\frac{P^0}{2}-\omega(q)-\omega(\vec{p}-\vec{q})}$
by $\frac{-1}{2M_\rho^2}$, which holds exactly at threshold with $\vec{q}=\vec{p}-\vec{q}=0$.

The explicit consideration of the propagator of the exchanged $\rho$ has produced a reduction factor in the loop with respect to its replacement by $1/(-M_{\rho}^2)$ as in Ref.~\cite{raquel}, but there are no singularities and no imaginary
part. In view of this, one can anticipate that one would get similar results using the approach of Ref.~\cite{raquel} but using an explicit cut-off that would effectively account for this converging factor. This means that if the cut-off is fine tuned to obtain the peak of $|T|^2$ at the mass of the $f_2(1270)$, one will need a smaller $q_\mathrm{max}$ in the approach of Ref.~\cite{raquel}  than explicitly using  the loop evaluated here, which is formally convergent. We will come back to this point later on.

Next we introduce the vertices. The right-hand vertex of Fig. \ref{fig:8} (b) is the contact term, $V_c$ of Eq. (\ref{eq:vs}). The two other vertices come from the combination
$(k_1+k_3)\cdot (k_2+k_4)=s-u$, where $k_1$, $k_2$ are the initial $\rho$ meson momenta and $k_3$, $k_4$ the outgoing ones. After projecting over the $s$-wave and taking the on-shell value, $k^2_i=m^2_\rho$, we obtain $-M^2_\rho V_{\mathrm{ex}}$, with $V_\mathrm{ex}$ of Eq. (\ref{eq:vs}). In principle, the $k_2$, $k_4$ momenta in the loop are off-shell. The on-shell factorization of this term (not of the $\rho$ exchanged propagator) is usually justified as follows \cite{npa,angels}.
We can write this term as $(s-u)_\mathrm{on} + [(s-u)_\mathrm{off}-(s-u)_\mathrm{on}]$. The $[(s-u)_\mathrm{off}-(s-u)_\mathrm{on}]$ can be written in powers of $(k^2_2-m^2_\rho)$ or $(k^2_4-m_\rho^2)$ (hence vanishing when $k^2_2=k^2_4=m^2_\rho$), and each of these terms kills one of the 
two intermediate $\rho$ propagators with momenta $q$ or $P-q$ in Fig. \ref{fig:9}. The remaining diagram (of the tadpole type if the $\rho$ propagator with three momenta $\vec{p}-\vec{q}$ in Fig. \ref{fig:9} is also shrunk) can usually be reabsorbed by the lowest order term, in a renormalization procedure. Following this philosophy we will also factorize the product of the two vertices with its on-shell value $-M^2_\rho V_\mathrm{ex}$. However, one cannot apply this procedure to the exchanged $\rho$ propagator because 
first, the intermediate $\rho$ states with momenta $q$ and $P-q$ in Fig. \ref{fig:9} cannot be placed on-shell for $\sqrt{s}$ below threshold; second, because as we have seen above, the exchanged $\rho$ with three momentum $\vec{p}-\vec{q}$ in Fig. \ref{fig:9} cannot be put on-shell; and third, because even if it could be placed on-shell, one still has the $d^3q$ integral to perform and the pole $(x-x_0+i\,\epsilon)^{-1}$ will give rise to ${\cal P} (x-x_0)^{-1}-i\,\pi\delta(x-x_0)$, both of them finite. Thus, the infinity which comes
from this propagator ``on-shell'' in Ref.~\cite{ollernew} is artificial. Our procedure, factorizing the vertices and keeping the full structure of the exchanged propagator, is a sensible one. The contribution of the diagram of Fig. \ref{fig:9} is then obtained, multiplying $t$ of Eq. (\ref{eq:eff}) by $V_c(-M^2_\rho)V_\mathrm{ex}$. The term of Fig. \ref{fig:8}(a) is obtained in the same way, substituting $-1/M^2_\rho$ in Eq. (\ref{eq:eff1}) by $V_c^2$. One may wonder what happens with higher order terms of the Bethe-Salpeter equation. One can see that all 
terms in this expansion, which do not have two consecutive $\rho$ exchanges, as in Fig. \ref{fig:8} (d), can be calculated without any difficulty. To make it technically easy we introduce an effective $\rho$ exchange propagator $G_{\rho,\mathrm{eff}}$ such that $G_{\rho,\mathrm{eff}}(s)G(s)=t(s)$, with $G$ the two $\rho$ meson loop function, $-M^2_\rho t_f$, of Eq. (\ref{eq:eff1}), and an effective $\tilde{V}_\mathrm{ex}$ potential
\begin{equation}
 \tilde{V}_\mathrm{ex}=V_\mathrm{ex}(-M^2_\rho)G_{\rho, \mathrm{eff}}.
\end{equation}
We see now that $(\tilde{V}_\mathrm{ex}+V_c)^2\,G$ gives rise by construction to the terms of Figs. \ref{fig:8}(a), (b), and (c), and provides an approximation for the term of Fig. \ref{fig:8} (d) as $\tilde{V}^2_\mathrm{ex} G$. With this approximation one gets the full Bethe-Salpeter series,
\begin{equation}
 T=[1-V_\mathrm{eff}G]^{-1} V_\mathrm{eff}
\end{equation}
with
\begin{equation}
V_\mathrm{eff}=\tilde{V}_\mathrm{ex}+V_c\ .
\label{eq:veff}
\end{equation}

Next we discuss the accuracy of 
the approximation done in the loop of Fig. \ref{fig:8}(d) with four $\rho$ meson propagators. This loop has been evaluated exactly in Appendix C of Ref.~\cite{geng}, where a lengthy expression is given. 
It has also been evaluated in
Ref.~\cite{raquel} with four pion propagators instead of $\rho$ propagators. Here
we can take advantage of the simplifications made in Eq. (\ref{eq:15.1}) and also evaluate 
exactly the loop with four meson propagators of Fig. \ref{fig:8}(d). This is done in the Appendix and the 
conclusion reached there is that the difference between the exact calculation
and $G^2_{\rho,\mathrm{eff}}G$ that we obtain using the effective potential ranges
from $18\%$ at $\sqrt{s}=1270$ MeV to $10\%$ at the $\rho\rho$ threshold. The approximation 
is acceptable when we know that the strength of this term is about
one fourth of the total one-loop contribution, which means we have 
$4.5\%$ difference in the total one-loop contribution at $\sqrt{s}=1270$ MeV and
$2.5\%$ difference at the $\rho\rho$ threshold. These differences are not relevant, even more when we know that small changes in a potential can be accommodated by small changes in the cut-off of $G$, which is finally fitted to the precise mass of a state.

There is one more point to discuss. The evaluation of $t$ in Eq.~(\ref{eq:15.1}) requires the knowledge of $\vec{p}$, the momentum of the initial $\rho$ 
in the molecule that is finally formed. Only the modulus is needed since $q$ is integrated over all angles and we can take $\vec{p}$ in the $z$ direction. In the ``on-shell'' factorization $\vec{p}\,^2$ was negative. Taking $\vec{p}\,^2$ negative is one way to say that one has negative energies with respect to the threshold, and in this sense it is used when one looks for poles of the $t$-matrix below threshold. However, in the physical systems the momenta are certainly real.  A bound state has negative energy
and a wave function which corresponds to a distribution of real momenta. A very good approximation to the wave functions derived with a potential of the type $V\theta(q_\mathrm{max}-q)\theta(q_\mathrm{max}-q')$, which leads to the standard Bethe Salpeter equation with a cutoff $q_\mathrm{max}$ in the
$G$ function\cite{gamer}, is given in Refs.~\cite{gamer,yamagata}. Using Eqs.~(105) of Ref.~\cite{yamagata} and Eq.~(47) of Ref.~\cite{gamer} we obtain
\begin{equation}
\langle p|\psi\rangle=g\frac{\theta(q_\mathrm{max}-p)}{E-\omega_1(p)-\omega_2(p)}
\end{equation}
where $g$ is the coupling of the state to the components of the wave function ($\rho\rho$ in this case). We determine an average momentum by looking at the peak of $p^2\langle p|\psi\rangle^2$ and we find $p\approx500 $ MeV/c for $E=1270$ MeV. This value could be smaller if the wave function picks up the lower components of the $\rho$ mass distribution, but we take this value for the evaluation, and $t$ is only smoothly dependent on $p$. For comparison, $p$ is of the order of 170 MeV/c for $E=1500$ MeV.


In Fig.~\ref{fig:11} we plot the effective potential $V_\mathrm{eff}$ of Eq.~(\ref{eq:veff}) as a function of the energy and compare it with the potential from Ref.~\cite{raquel} and from the ``on-shell'' factorized potential, already shown in Fig.~\ref{fig:3}. As we can see, $V_\mathrm{eff}$ is smaller than the potential from Ref.~\cite{raquel}, which is logical since it incorporates the $q^2$ dependence of the $\rho$ propagator. Yet, the potential does not have any singularity, as is the case of $V(s)$, and we showed that the propagator in the loops does not develop a singularity.  Also, $V_\mathrm{eff}$ below threshold does not have an imaginary part, unlike $V(s)$ which develops an imaginary part with a discontinuity at $s=3M_\rho^2$.

In Fig. ~\ref{fig:12} we show the results for $|T|^2$ using $V_\mathrm{eff}$. As anticipated, in order to have a bound state at 1270 MeV, we must use a larger value of $q_\mathrm{max}$ than in the case of the potential in Ref.~\cite{raquel} because $V_\mathrm{eff}$ already includes the effects of $q^2$ in the $\rho$ propagator, which reduces
the contributions of the $\rho$ exchange potential. Such effects are effectively taken into account in Ref.~\cite{raquel} by using a smaller cut-off $q_\mathrm{max}$. The calculation of Fig.~\ref{fig:12} is done, as in Fig.~\ref{fig:7}, using the convoluted $\tilde{G}$ function to account for the mass distribution of the $\rho$.  The use of the convoluted $\tilde{G}$ function in the Bethe-Salpeter equation gives a width to the state because it can now decay to $\rho\pi\pi$ or $\pi\pi\pi\pi$. We already mentioned that this provides only part of the width. In the case of the $f_2(1270)$ most of the width comes from $\pi\pi$ decay, which we evaluated in Ref.~\cite{raquel} by means of a box diagram. We refrain from doing it here, but the small width obtained using $V_\mathrm{eff}$ or the potential of Ref.~\cite{raquel} serves us the purpose of evaluating the coupling of the state to $\rho\rho$, which we do in the following way \cite{nagahiro}:
\begin{equation}
g_T^2=M_R\Gamma_R \sqrt{|T|^2_\mathrm{max}}
\end{equation}
where $M_R$, $\Gamma_R$ are the mass and width respectively of the tensor state in Figs.~\ref{fig:7} and \ref{fig:12}, and $|T|^2_\mathrm{max}$ is the value of $|T|^2$ at the peak. The value of $g_T$  is $g_T=10700\,\mathrm{MeV}$
in the calculation with the effective potential and a cut-off of $1500$ MeV, and $
g_T=11700\,\mathrm{MeV}$
with the potential of Ref. \cite{raquel}  and a cut-off of $860$ MeV. In both cases,  the pole shows up at $\sqrt{s_0}=1273$ MeV with a width of  $3$ MeV. If $p=50$ MeV, the pole with the effective potential appears at $\sqrt{s_0}=1254$ MeV, with $\Gamma=2$ MeV and $g_T=10000$ MeV, while if $p=800$ MeV, we obtain the pole at $\sqrt{s_0}=1300$ MeV, with $\Gamma=5$ MeV and $g_T=11000$ MeV.

The value of $g_T$ is very similar in both approaches, with differences of less than 10\%. This connects with our discussion in the Introduction because the compositeness condition \cite{weinberg, kalash,gamer} provides the coupling as a function of the binding energy for small binding, and in the present case, the fact that $g_T$ is roughly model independent is somehow telling us that the wave function has picked up the low mass components of the $\rho$, which provide less binding. 

The  fact that $g_T$ is so stable and the results obtained are so close to those obtained before with the extreme approximation of neglecting the $q^2$ dependence of the $\rho$ propagator, but coping for it by means of a reduced cut off, is very important and reinforces the agreement found with the couplings of Ref.~\cite{raquel} for the radiative decay of this resonance \cite{nagahiro} and other decays \cite{minirev}.

\section{Conclusions}
We have made a critical discussion of a recent work \cite{ollernew} where certain improvements have been made in the $\rho\rho$ interaction. Yet, the use of the Bethe-Salpeter equation with an ``on-shell'' factorization of the potential leads the authors to conclude that, unlike the $f_0(1370)$ state, which appears as a bound $\rho\rho$ state, the tensor state $f_2(1270)$, which had been obtained before with some non-relativistic approximations, disappears. We argue from a general point of view that if the potential for $J=2$ is more than twice more attractive than the case of $J=0$ (as is the case in Ref.~\cite{ollernew}) and the $J=0$ bound state is found in Ref.~\cite{ollernew}, the appearance of a bound state in $J=2$ is unavoidable. Then we proceed to understand the reason for the claim in Ref.~\cite{ollernew}. The problem stems from the ``on-shell'' factorization of the potential on top of a singularity which produces a ``potential'' of infinite strength and with a big imaginary part that has a discontinuity in the singular point. We show that this imaginary part is unphysical and bears no connection to the decay products of the $\rho\rho$ bound state into $\pi\pi$ or $\pi\pi\pi\pi$. 
After the source of the anomalous results in Ref.~\cite{ollernew} is disclosed, we proceed to tackle the problem in an appropriate way, evaluating the loops with the full $\rho$ propagators for the $\rho$ in the exchange channel, and see that there are no singularities nor an imaginary part below threshold tied to those diagrams. Finally, from the evaluated loops we define an effective potential in a way that, when used with the Bethe-Salpeter equation, renders the results of the loop. With this effective potential we solve the Bethe-Salpeter equation and find a bound state for $J=2$. Upon fine tuning of the cut-off in the $G$ function, taking into account the $\rho$ mass distribution, the bound state is made to appear at 1270 MeV to generate the $f_2(1270)$ resonance, and its couplings to the $\rho\rho$ component are extracted. Then we find that the coupling evaluated with this improved method is very similar to the one obtained with a more drastic approximation made in Ref.~\cite{raquel}, where in analogy to the construction of the chiral Lagrangians starting from the local hidden gauge Lagrangians, the $q^2$ in the propagators of the exchanged vector mesons is removed. We show that after tuning the cut-off with this latter approximation, to approximately take into account the reduction of the exchanged propagators due to their $q^2$ dependence, and fitting the energy of the bound state to the experimental one, the resulting $T$ matrix around the bound state energy is remarkably similar to  the one obtained with the more sophisticated approach of the effective potential. 
\section{Appendix}
\subsection{Comparing the loop with four $\rho$ mesons with the loop with $D_{\rho,\mathrm{eff}}$}
We can easily compare these two magnitudes starting from Eq. (\ref{eq:15.1}). 
The introduction of an extra
$\rho$ exchange propagator, $\left[(\frac{P_0}{2}-q^0)^2-(\vec{p}-\vec{q})^2-M^2_\rho\right]^{-1}$, where we take $\vec{p}$ and $\vec{p}\,'$ ( the three momentum of the outgoing $\rho$) as equal for simplicity, can be obtained  by changing $M^2_\rho\to M^{'\,2}_\rho$ in this propagator in Eq. (\ref{eq:15.1}), which we will call $t'$, and evaluating 
\begin{equation}
\frac{\partial t'}{\partial M_\rho^{'\,2}}.
\label{eq:m1}
\end{equation}
We then compare this with
\begin{eqnarray}
 G^2_{\rho,\mathrm{eff}}G\equiv \left(\frac{t}{G}\right)^2G=\frac{t^2}{G}=\frac{t^2}{-M^2_\rho\, t_f}\ .
\label{eq:m2}
 \end{eqnarray}
The two magnitudes, Eq. (\ref{eq:m1}) and Eq. (\ref{eq:m2}) differ by $18\%$ at $\sqrt{s}=1270$ MeV, and the 
difference goes down to $10\%$ at the $\rho\rho$ threshold.

\section*{Acknowledgments}
We acknowledge some discussions with J.A. Oller and U. -G. Mei\ss ner.
This work is partly supported by the National Natural Science Foundation
of China (Grants No.11375024 and  No.11522539). This work is also partly supported by the Spanish
Ministerio de Economia y Competitividad and European FEDER funds
under the contract number FIS2011-28853-C02-01, FIS2011-
28853-C02-02, FIS2014-57026-REDT, FIS2014-51948-C2- 1-P, and
FIS2014-51948-C2-2-P, and the Generalitat Valenciana in the program
Prometeo II-2014/068. We acknowledge the support of the European
Community-Research Infrastructure Integrating Activity Study of
Strongly Interacting Matter (acronym HadronPhysics3, Grant Agreement
n. 283286) under the Seventh Framework Programme of the EU.

  \clearpage
  \end{CJK*}
\end{document}